
%
%
%
%
\input harvmac.tex

\def\frac#1#2{{\textstyle{#1\over #2}}}
\def\nl{nonlocal~}
\def\cl{conservation law~}
\def\cls{conservation laws~}

\overfullrule=0pt

\newcount\eqnum
\eqnum=0
\def\eq{\eqno(\secsym\the\meqno)\global\advance\meqno by1}
\def\eqlabel#1{{\xdef#1{\secsym\the\meqno}}\eq }

\newwrite\refs
\def\startreferences{
 \immediate\openout\refs=references
 \immediate\write\refs{\baselineskip=14pt \parindent=16pt \parskip=2pt}
}
\startreferences

\refno=0
\def\aref#1{\global\advance\refno by1
 \immediate\write\refs{\noexpand\item{\the\refno.}#1\hfil\par}}
\def\ref#1{\aref{#1}\the\refno}
\def\refname#1{\xdef#1{\the\refno}}

\def\pl#1#2#3{{\it Phys. Lett.} {\bf B#1#2#3}}

\newcount\exno
\exno=0
\def\Ex{\global\advance\exno by1{\noindent\sl Example \the\exno:

\nobreak\par\nobreak}}

\parskip=6pt
\Title{\vbox{\baselineskip12pt
\hbox{LAVAL-PHY-21/93}}}
{\vbox {\centerline{Nonlocal conservation laws for supersymmetric KdV
equations} }}
\centerline{P. Dargis\foot{Work supported by NSERC
(Canada).} and P. Mathieu\foot{Work supported
by NSERC (Canada) and FCAR (Qu\'ebec).} }
\vskip.2in
\smallskip\centerline{\it D\'epartement de Physique,
Universit\'e Laval, Qu\'ebec, Canada G1K 7P4}
\vskip .2in
\centerline{\bf Abstract}
\bigskip
\noindent
The \nl \cls for the N=1 supersymmetric KdV equation are shown to be related in
a simple way to powers of the fourth root of its Lax operator.  This provides
a direct link between the supersymmetry invariance and the existence of \nl
conservation laws.  It is also shown that nonlocal conservation laws exist for
the two integrable N=2 supersymmetric KdV equations whose recursion operator is
known.

\Date{1/93\ \ (hepth@xxx/9301080)}


\newsec{Introduction}

The supersymmetric KdV (sKdV) equation [\ref{Yu.I. Manin and A.O. Radul,
{\it Comm. Math. Phys.} {\bf 98} (1985) 65}\refname\yu,\ref{P. Mathieu, {\it
J. Math. Phys.} {\bf 29} (1988) 2499}\refname\mjmp] has been much studied in
the
last few years [\ref{I. Yamanaka and R. Sasaki, {\it Prog. Theor. Phys.} {\bf
79}
(1988) 1167}\refname\yam\aref{P. Mathieu, {\pl203} (1988)
287}\refname\mat\aref{T. Inami and H. Kanno, {\it Comm. Math. Phys.} {\bf 136}
(1991) 519}\refname\ina\aref{W. Oevel and Z. Popowicz, {\it Comm. Math. Phys.}
{\bf 139} (1991) 441 }\refname\oev \aref{J.M. Figuera-O'Farrill, J. Mas, and
E. Ramos, {\it Leuven preprint} KUL-TF-91-19}\refname\fig\aref{I.N. McArthur,
{\it Comm. Math. Phys.} {\bf 148} (1992) 177}\aref{P. Mathieu, in ``Integrable
and Superintegrable Systems", ed. B. Kupershmidt (World Scientific,
1990) p352}\aref{E.D.van der Lende and H.G.J.Pijls, {\it Indag. Math.} N.S.
{\bf
1:2}, (1990) {221}; E.D.van der Lende, PhD
thesis}\refname\van-\ref{P.H.M. Kersten, {\it Phys. Lett.} {\bf 134 A} (1988)
25}\refname\kers].  Its hamiltonian forms have been obtained in two steps. The
second hamiltonian structure, related to the superconformal algebra
[\yam,\mat],
has been obtained some time ago [\mjmp].  However it is only recently that its
first hamiltonian structure has been discovered [\oev,\fig] and it turns out to
be nonlocal.\foot{The adjectives `first' and `second' have thus no
chronological
origin.  They are inherited from their KdV relatives.}  Both structures
were shown to be related to the Lax operator by the standard Gelfand-Dickey
brackets after reduction [\oev,\fig]. With these two hamiltonian structures,
one
can construct a recursion operator from which the infinite sequence of
conservation laws can be obtained once the simplest one is known.  These
conservation laws can also be obtained in the standard way, from the
super-residue of fractional powers of the Lax operators [\yu,\mjmp].

There remains a fact that has not received a clear explanation,
namely the occurrence of nonlocal conservation laws. They have been found in
[\kers] from a tedious symmetry analysis and traced back to the mere existence
of the supersymmetry invariance. Here we show that these \nl \cls are related
to
the Lax operator. We also point out that, with a natural extension of the
variationnal calculus, they can be generated recursively. Finally these results
are partly extended to N=2 supersymmetric KdV (SKdV - where capital S stands
for
N=2) equations. For the two SKdV equations whose Lax operator is known
[\ref{C.A. Laberge and P. Mathieu, {\it Phys. Lett.} {\bf B215} (1988)
718}\refname\laber] we prove, in each case, that there are two infinite
sequences of \nl conservation laws which can be generated recursively.  This
was
an expected result given the existence of two supersymmetries.  However their
relation to the Lax operators still has to be clarified.  On the other hand,
there is a third SKdV equation expected to be integrable [\ref{P. Labelle and
P. Mathieu, {\it J. Math. Phys.} {\bf 32} (1991) 923}\refname\label], for which
no nontrivial \nl \cls have been found.

\newsec{N=1 sKdV equation}

Let us introduce the fermionic superfield (with an implicit time dependence)
$$\phi(x,\theta)=\theta u(x)+\xi(x)\eqlabel\compphi$$
where $\theta$ is a
Grassmann space variable and $u(x)$ is the usual KdV field. In term of this
superfield and the superderivative D defined as
$$\eqalign{D^{~}\rm&=\mit \theta \rm \partial +{\partial }_{\mit \theta }\cr
{D}^{\rm 2}&=\partial \equiv {\partial }_{\mit x}\cr}\eqlabel\sddef$$
the sKdV equation reads [\mjmp]
$$\phi_t=-\phi_{xxx}+3(\phi D\phi)_x\eqlabel\skdv$$
Its Lax operator is [\yu,\mjmp]
$$L={\partial}^2-\phi D~~{\rm or}~~{\partial}^2+\phi D - (D\phi)\eq$$
(Parentheses are used to delimit the action
of the derivatives).  The degeneracy of the Lax formalism has been explained
Lie
algebraically in [\ina].  For calculations one can use either form.  In terms
of
$L$, the usual local \cls can be written as
$$H_n=\int dX~Res~L^{n/2}\eqlabel\claw$$
$dX$ stands for $dxd\theta$ ( $\int d\theta=0~{\rm and}~\int d\theta~\theta=1$
)
while the residue ($Res$) of a super-pseudodifferential operator is the
coefficient of $D^{-1}$.  The explicit form of the first few can be found
in [\mjmp,\yam].  Given that $L^{n/2}={\partial}^n+\ldots$ is bosonic, its
residue is
fermionic.  The measure being also fermionic, these \cls are bosonic.  The two
hamiltonian operators are [\mjmp,\oev,\fig]
$$P_1=\partial[D^3-{\phi}]^{-1}\partial\eq$$
$$P_2=-D^5+3\phi\partial+(D\phi)D+2{\phi}_x\eq$$
from which one constructs the recursion operator
$$R=P_1^{-1}P_2 = ( D^{-1} -
{\partial}^{-1}\phi{\partial}^{-1})P_2\eqlabel\recur$$

Let us now turn to the \nl \cls.  The first few of them can be easily obtained
from (\skdv) (i.e., one writes down the most general \nl differential
polynomial of appropriate degree, with $deg(\phi)={\rm 3/2},~deg(D)={\rm 1/2}$,
 and adjusts its coefficients such that its integral is time independent).
These
are
$$\eqalign{
&{J}_{1/2}=\int dX \left({{D}^{-1}\phi}\right) \cr
&{J}_{3/2}=\int dX {\left({{D}^{-1}\phi }\right)}^{2} \cr
&{J}_{5/2}=\int dX \left[{ {\left({{D}^{-1}\phi}\right)}^{3}-6\left({D}^{-1}
                         \left({\phi D\phi }\right)\right)  }\right] \cr
&{J}_{7/2}=\int dX \left[{ {\left({{D}^{-1}\phi}\right)}^{3}-12{\left({{D}\phi
}\right)}^{2}-24\left(D^{-1}\phi\right)\left(D^{-1}\left(\phi
D\phi\right)\right)   }\right]
}\eqlabel\a$$
The subscripts indicate the degree.  One first
notices that they are all fermionic and that there is one \cl at each half
integer degree.  ( Notice that the $H_n$ are non-zero only for n odd, hence
there are twice as many \nl \cls than local ones.)   The infinite sequence can
be constructed from ${J}_{\rm 1/2}$ and  ${J}_{\rm 3/2}$ by using the recursion
operator :
$$R{{\delta}\over{\delta\phi}}{J_{1/2+i+2n}}=
{{\delta}\over{\delta\phi}}{J_{1/2+i+2 n+2}}\qquad n=0,1,2,...~~i=0,1\eq$$
For this, however, one has to extend the validity of the usual formula for the
variational derivative (recall that here $\phi$ is odd)
[\yu,\van,\ref{P. Mathieu,  {\it Lett. Math. Phys.} {\bf 16} (1988) 199}]
$${\delta  \over \delta \phi }\int_{}^{}dX~\rm \mit h\rm (\mit \phi \rm )=
\sum\nolimits\limits_{\mit k}^{} {\rm (-)}^{\mit k\rm +\mit k\rm (\mit k\rm
+1)/2}\left({\mit{{D}^{k}{\partial h \over \partial ({D}^{k}\phi
)}}}\right)\eq$$
to negative values of $k$.  For instance
$${{\delta}\over{\delta\phi}}{\int dX~(D^{-1}\phi)}=
-(D^{-1}1)=-\theta\eqlabel\quinze$$
(since $(D\theta)=1$). Similarly, the validity of the formula for integration
by
parts,
$$\int{dX~(D^k A)B}=(-)^{k\tilde{A} +k(k+1)/2}\int{dX~A(D^k
B)}\eq$$ where $\tilde{A}=0~(1)~~{\rm if}~~A$ is even (odd), is extended to the
case where $k<0$.  In particular one has
$$\int{dX~({D}^{-1}\phi)}=
-\int{dX~\phi({D}^{-1}1)}=-\int{dX~\phi\theta}  \eq$$
from which one can recover the result (\quinze).  Notice that with this
definition, $u(\partial^{-1}u)$ is a total derivative, i.e.,
$\int{dx~u(\partial^{-1}u)}=-\int{dx~(\partial^{-1}u)u}=0$.

\noindent{{\it Example}: Here are few steps of a sample calculation leading
from
$J_{3/2}$ to $J_{7/2}$ :}
$$\eqalign{R{1\over2}{{\delta}\over{\delta\phi}}{J_{3/2}}
&=R(-(\partial^{-1}\phi))\cr
=&-(D^{-1} - \partial^{-1}\phi\partial^{-1})\partial
   (-(D\phi)+2\phi(\partial^{-1}\phi)+(D^{-1}(\phi(D^{-1}\phi))))\cr
=&\phi_x - 2(D\phi)(\partial^{-1}\phi)+\phi(D^{-1}\phi)
  - (\partial^{-1}(\phi(D\phi)))+{\frac16}(D^{-1}(D^{-1}\phi)^3) \cr
=&-{1\over{24}}{\delta  \over \delta \phi }\int{dX~[(D^{-1}\phi)^4
  - 12(D\phi)^2 - 24(D^{-1}\phi)D^{-1}(\phi D \phi)]}}\eq$$
To get the second equality, we used the relations
$$(D\phi)(D^{-1}\phi)=(D(\phi D^{-1}\phi))~~,~~(\phi_x \partial^{-1}\phi)=
(\partial(\phi \partial^{-1}\phi ))\eq$$
which holds because $\phi^2=0$, $\phi$ being fermionic.  To get the third
equality we used
$$(D^{-1}(\phi D^{-1}\phi))=\frac12(D^{-1}\phi)^2\eq$$
This is easily checked by setting $\phi=DF$ ($F$ being thus bosonic):
$$(D^{-1}(\phi D^{-1}\phi)) = (D^{-1}(FDF)) = \frac12(D^{-1}D(F^2))
=\frac12(D^{-1}\phi)^2\eq$$
Notice that ${{\delta}\over{\delta\phi}}{J_{1/2+2 n}}$ always
contains explicit $\theta$ terms while ${{\delta}\over{\delta\phi}}{J_{3/2+2
n}}$
does not.

Thus, given $J_{1/2}$, $J_{3/2}$ and $R$, we conclude that there exists an
infinite sequence of odd \nl conservation laws.  But is there a more
direct way to probe their existence ?  The clue to the answer is to notice that
$L$ admits not only a square root but also an odd fourth root, of the form
$L^{1/4}=D+...$.  It is then clear that (up to a constant) the fermionic \cls
must be related to $L$ via
$$J_{l/2}=\int dX~Res~L^{l/4}\qquad(l~\rm odd)\eqlabel\holmes$$
Their nonlocality is rooted in the nonlocality of $L^{1/4}$ itself, i.e.
$$L^{1/4}=D+(\partial^{-1}\phi)-\frac12(D^{-1}\phi)D^{-1}-\frac12\phi D^{-2}
+\frac14(D\phi)D^{-3}+\frac18(D^{-1}\phi)^2D^{-3}+...\eq$$
Now all these odd \nl \cls commute with the usual local bosonic conservation
laws.  This follows trivially from the fact that the latter are possible
hamiltonians and the former are time independent.  However, what about the
commutation of the fermionic \cls among themselves?  From the Jacobi identity
for the Poisson brackets of $J_{l/2}$, $J_{k/2}$ and $H_n$, it follows that
$\{J_{l/2},J_{k/2}\}$, calculated with the second Poisson structure, is
necessarily a \cl, and its degree is ${{l+k}\over2}$.  Therefore up to
numerical
factors, one must have
$$\eqalign{ &\{J_{(4 n+1)/2},J_{(4 m+1)/2}\}=H_{2(n+m)+1} \cr
&\{J_{(4n+3)/2},J_{(4 m+3)/2}\}=H_{2(n+m)+3} \cr
&\{J_{(4 n+1)/2},J_{(4 m+3)/2}\}=0
}\eqlabel\e$$
The last result is a consequence of the absence of \cls with even integer
degree.  The first two are readily verified explicitly by using the first few
\cls given in (\a).  One checks in particular that the proportionality
constants are nonzero.  This algebra shows that \nl \cls are some sort of
square root, in a Poisson bracket sense, of local conservation laws.

Let us now prove these relations at the level of Lax equations.  For this
we introduce a new infinite sequence of odd {flows}
$$\partial_{\tau_l}L=[G^l_+,L]\eq$$
generated by the fourth root of $L$:
$$G\equiv L^{1/4}\eq$$
$l$ is restricted to be an odd integer so that the $\tau_l$ are odd
parameters.  These imply directly that
$$\partial_{\tau_l}G^k=[G^l_+,G^k]\eqlabel\star$$
where the commutator is
understood as a graded commutator, i.e.
$$[A,B]=AB-(-)^{\tilde A \cdot \tilde B}BA\eq$$
A direct calculation yields
$$(\partial_{\tau_l}\partial_{\tau_k}+\partial_{\tau_k}\partial_{\tau_l})L=
[{\cal B},L]\eq$$
where ${\cal B}$ is given by
$${\cal B}=[G^k_+,G^l]_++[G^l_+,G^k]_+-[G^l_+,G^k_+]\eq$$
To obtain this equality, we used the graded Jacobi identity
$$[[A,B]C]+[[C,A]B](-)^{\tilde C (\tilde A + \tilde B)}
+[[B,C]A](-)^{\tilde B (\tilde C + \tilde A)}=0\eq$$
Now with $G^k=G^k_++G^k_-$ and $[G^k_-,G^l_-]_+=0$, one can rewrite ${\cal B}$
as
$${\cal B}=(G^kG^l+G^lG^k)_+=2L_+^{{l+k}\over2}\eq$$
This shows that
$$\partial_{\tau_k}\partial_{\tau_l}+\partial_{\tau_l}\partial_{\tau_k}=
2\partial_{t_{k+l}}\eqlabel\sstar$$
where the flow $t_n$ is defined as
$$\partial_{t_n}L=[L^{n/2}_+,L]\eq$$
The flow $t_{k+l}$ is trivial if $k+l$ is even, and generated by $H_{k+l}$ if
$k+l$ is odd.  This completes the proof of (\e).

As a simple check of (\sstar), let us verify that
$\partial^2_{\tau_{3/2}}=\partial_{t_3}$.  With
$$L^{3/4}_+=D^3+(\partial^{-1}\phi)\partial-\frac12(D^{-1}\phi)D\eq$$
one easily finds that
$$\partial_{\tau_{3/2}}\phi=D^3\phi+(\partial^{-1}\phi)(\partial \phi)
-\frac12(D^{-1}\phi)(D\phi)\eqlabel\waynesworld$$
Applying $\partial_{\tau_{3/2}}$ on this equation gives
$$\partial^2_{\tau_{3/2}}\phi=+\frac14[-\phi_{xxx}+3(\phi D\phi)_x]\eq$$
(\waynesworld) is thus the square root of the sKdV equation.  On the
other hand, we have checked directly that the first three \nl \cls are indeed
given by (\holmes).  In particular, deriving $J_{5/2}$ in this way is
non-trivial (and very long) since it contains two terms, hence a precise
relative coefficient.

Relations (\e) and (\sstar) can be regarded as the global translation of
the graded commutation relations obtained in [\kers] from the symmetry
algebra.

A final remark will close this section.  The \nl conservation laws,
being written as a superintegral of a density expressed in terms of the
superfield $\phi$ and the superderivative $D$, appear to be manifestly
supersymmetric invariant.  However the nonlocality may induce a breaking of
supersymmetry.  In fact all $J_{(4n+1)/2}$ are not invariant under a
supersymmetry transformation.  This is most simply seen with the
component formulation (\compphi).  The supersymmetric transformation of the
fields is
$$\delta u=\eta \xi_x\qquad \delta \xi = \eta u\eq$$
where $\eta$ is a constant anticommuting parameter.  In components, $J_{1/2}$
reads $\int dx~\xi$ so that
$$\delta J_{1/2} = \eta\int dx~u\neq 0\eq$$
On the other hand, for $J_{3/2}$ one has
$$\delta J_{3/2} = \delta\int dx~\xi (\partial^{-1}u)
= \eta\int dx~(u(\partial^{-1}u)-\xi\xi)= 0\eq$$
(the first term in the integrand being a total derivative), that is $J_{3/2}$
is
invariant under a supersymmetry transformation.  The following observation
shows
that $\delta J_{(4n+3)/2}=0$ simply because $H_{2n+2}=0$.  Indeed, up to a
constant factor, the action of $\delta$ is equivalent to taking the Poisson
bracket of the second hamiltonian structure with $J_{1/2}$, e.g.
$$\delta\cdot=-\eta\{J_{1/2},\cdot\}\eq$$
With this, the above results are seen to be simple transcriptions of the
first and third relations in (\e) for $n=0$.

\newsec{N=2 SKdV equations}

We now introduce an even N=2 superfield $\Phi(x,\theta_1,\theta_2)$ whose
components are
$$\Phi(x,\theta_1,\theta_2)=\theta_2\theta_1 u(x)
+\theta_1\xi_1(x)+\theta_2\xi_2(x)+w(x)\eq$$
Here $\theta_{1,2}$ are two anticommuting variables, $\theta_1^2=\theta_2^2=0,~
\theta_1\theta_2=-\theta_2\theta_1$, $\xi_{1,2}$ are two anticommuting fields
and $w$ is a new bosonic field.  Notice that $deg(u,\xi_i,w)=(2,3/2,1)$.  One
also introduces two superderivatives $D_{1,2}$ defined as
$$\eqalign{&D_i=\theta_i\partial+\partial_{\theta_i} \cr
           &D_1^2=D_2^2=\partial\qquad D_1D_2=-D_2D_1}\eq$$
The two integrable SKdV equations, whose Lax operators are known are [\laber]
$$\eqalign{{\Phi }_{t} &=-{\Phi }_{xxx}
+3{\left({\Phi {D}_{1}{ D}_{2}\Phi }\right)}_{x} +{1
\over 2}\left({\alpha  -1}\right){\left({{\mit D}_{1}{D}_{
2}{\Phi }^{2}}\right)}_{x}\rm +3 \alpha {\Phi }^{2}{\Phi }_{x} \cr
&\alpha=-2,4}\eqlabel\SKDV$$
and denoted by ${\rm SKdV_{-2}~and~SKdV_4}$.  The Lie algebraic structure
underlying the first system was unravelled in [\ref{T. Inami and
H. Kanno, {\it Nucl. Phys.} {\bf B359} (1991) 201}].  The corresponding Lax
operators are
$$\eqalign{
L_{(-2)}&:\partial^2+2\Phi D_1D_2-(D_2\Phi)D_1+(D_1\Phi)D_2 \cr
L_{(4)}~&:\partial^2 -2\Phi D_1D_2 +(D_2\Phi)D_1-(D_1\Phi)D_2
-(D_1D_2\Phi)-\Phi^2 \cr
&~=-(D_1D_2+\Phi)^2
}\eq$$
and the two are self-adjoint.  Both systems are bi-hamiltonian.  They share a
common second hamiltonian structure [\laber]
$$\hat P_2=D_1D_2\partial+2\Phi\partial-(D_1\Phi)D_1-(D_2\Phi)D_2
+2\Phi_x\eq$$
and their first hamiltonian structure is given respectively by [\oev]
$$\eqalign{\hat{P}_{1~(-2)}&=(D_1D_2\partial^{-1}-D_1^{-1}\Phi D_1^{-1}
-D_2^{-1}\Phi D_2^{-1})^{-1} \cr
\hat{P}_{1~(4)}~&=\partial}\eq$$
One then has a recursion operator $\hat R =\hat P^{-1}_1\hat P_2$ for each
hierarchy.  Notice however that $degR_{(4)}=1$ while $degR_{(-2)}=2$.  This
translates into the fact that there are twice as many local \cls for
 the ${\rm SKdV_{4}}$ equation as for the ${\rm SKdV_{-2}}$ one.  There exists
one local \cl for each integer degree in the former case and one for each odd
integer degree in the latter one.  Conservation laws for odd degrees in both
cases are related to the Lax operator by
$$H_n=\int{d\hat{X}~\hat{R}es~L^{n/2}}\qquad n{\rm~odd}\eq$$
where the N=2 super residue is defined as
$$\hat Res={\rm coeff~of~} D_1D_2\partial^{-1}\eq$$
$d\hat X = dxd\theta_1 d\theta_2$ and $L^{n/2}=\partial^n+...$.  The origin
of the \cls of even degree for the case $\alpha=4$ is rooted in the existence
of
another square root for $L_{(4)}$ (different from $\partial+...$) namely [\oev]
$$L'=D_1D_2+\Phi\qquad(L')^2=-L_{(4)}\eq$$
In term of $L$ and $L'$ one can introduce the new flows [\oev]
$$\partial_{t_n'}L=[(L^{n/2}L')_+,L]\eq$$
giving rise to another infinite sequence of \cls.
$$H_{n+1}=\int{d\hat{X}~\hat{R}es~(L^{n/2}L')}\qquad n{\rm~odd}\eq$$

For these N=2 supersymmetric KdV equations, there exists also \nl
conservation laws.  The first few of them are

\noindent{$\alpha=-2$}
$$\eqalign{
&J_{1/2}^{(1)}=\int{d\hat X~ (D_1^{-1}\Phi)} \cr
&J_{3/2}^{(1)}=\int{d\hat X~ \Phi(D_1^{-1}\Phi)} \cr
&J_{5/2}^{(1)}=\int{d\hat X~ [-{\frac23}(D_1^{-1}\Phi^3)
+{\frac16}\Phi^2(D_1^{-1}\Phi)+(D_1^{-1}(\Phi D_1D_2\Phi))} \cr
&\qquad\quad -{\frac16}(D_1^{-1}\Phi)(D_1D_2\partial^{-1}\Phi)^2 ]
 }\eqlabel\alpa$$

\noindent{$\alpha=4$}
$$\eqalign{
&J_{1/2}^{(1)}=\int{d\hat X~ (D_1^{-1}\Phi)} \cr
&J_{3/2}^{(1)}=\int{d\hat X~ [\Phi(D_1^{-1}\Phi) -2(D_1^{-1}\Phi^2)]} \cr
&J_{5/2}^{(1)}=\int{d\hat X~ [{\frac43}(D_1^{-1}\Phi^3)
-{\frac56}\Phi^2(D_1^{-1}\Phi)+(D_1^{-1}(\Phi D_1D_2\Phi))} \cr
&\qquad\quad -{\frac16}(D_1^{-1}\Phi)(D_1D_2\partial^{-1}\Phi)^2 ]
 }\eqlabel\alpb$$
In each case there is another infinite series $J_{l/2}^{(2)}$ obtained from
$J_{l/2}^{(1)}$ with $D_1\leftrightarrow D_2$.  The first two fermionic \cls
for
$\alpha =-2$ in (\alpa) are easily obtained by inspection; from these two, all
others can be generated by the action of the recursion operator.  On the other
hand for $\alpha =4$, one only needs to know $J_{1/2}^{(1)}$ to get all the
$J_{l/2}^{(1)}$'s recursively.

Computing the Poisson brackets of these first few \nl conservation laws, we
infer
the algebraic structure
$$\{J_{l/2}^{(1)},J_{k/2}^{(2)}\}=0\eq$$
for any $l$ and $n$, and
$$ \{ J_{l/2}^{(i)},J_{k/2}^{(i)} \} =
\left \{ \eqalign{ H_{l+k}\qquad\quad &{\rm for~}{{l+k}\over2}{\rm ~odd} \cr
\delta_{\alpha,4}H_{l+k}\qquad &{\rm for~}{{l+k}\over2}{\rm ~even}
} \right.\eq$$
with $i=1{\rm ~or~}2$.

Unfortunately, the Lax origin of these \nl \cls has not beeen unraveled.  The
operator $L$ would seem to have two distinct fourth roots:
$$\eqalign{
&G_{(i)}=D^i+{\rm lower~order~operators} \cr
&G_{(i)}^4=L\quad i=1,2
}\eq$$
whose residues would be natural candidates for the \nl conserved densities.
But a simple calculation shows that such $G_{(i)}$ do not exist.  A variant of
this idea, where $L^{1/2}$ is factorized into a symmetric product of two odd
operators, is not succesful either.

On the other hand, it has been conjectured
in [\label] that the ${\rm SKdV_{1}}$ equation is also integrable on the basis
of the existence of $H_5$ and $H_7$ (yet two other \cls have been found
[\ref{C.M. Yung, {\it Tasmania preprint}, UTAS-PHYS-92-17}]).  It is natural to
see whether it has nonlocal \cls.  Of course $J_{1/2}^{(1,2)}$ is conserved.
However it is not difficult to check that there are no \nl fermionic \cls of
the
general form
$$J_{3/2}^{(1)}(\alpha)=\int{d\hat X~[a\Phi(D_1^{-1}\Phi)
+b(D_1^{-1}\Phi^2)]}\eq$$
for values of $\alpha$  different from $4$ and $-2$.

\newsec{Concluding remarks}

In this work we have shown that \nl \cls of supersymmetric KdV equations can be
analysed with essentially the same methods used for local ones.  Actually, they
can be regarded as some sort of square root (in a Poisson brackets sense) of
the
usual bosonic local conservation laws, an interpretation which clearly reflects
the supersymmetric invariance.

It is clear that the existence of \nl\cls will modify the algebraic structure
of mastersymmetries obtained in [\oev] for N=1.  Most probably it will change
the
centerless Virasoro algebra into its appropriate supersymmetric extension.

On the other hand, the local \cls for the quantum version of the system
considered here appear in the appropriate perturbed superconformal minimal
models
[\ref{P. Mathieu, {\it Nucl. Phys.} {\bf B336} (1990) 338; P. Mathieu and
M. Walton, {\it Phys. Lett.} {\bf 254} (1991) 106}].  It is natural to expect
that the quantum form of the \nl\cls will also be present in these
off-critical theories.  Actually it is rather simple to check this for the
first
few \nl\cls of the system considered.  But these are conservation laws which
contain a single term.  In the quantum case it turns out to be difficult to
calculate the relative coefficients of the different terms of a \nl
conservation
law. We expect to return to this elsewhere.  In that context, it should be
stressed that the \nl \cls considered here have no relation with those found in
[\ref{D. Bernard and A. LeClair, {\it Comm. Math. Phys.} {\bf 142} (1989) 99 ;
C. Ahn, D. Bernard and A. LeClair, {\it Nucl. Phys.} {\bf B346} (1990) 409}]
for
perturbed conformal field theories.  At first the latter have a purely quantal
origin, and furthermore they cannot be expressed in terms of the
energy-momentum
tensor, except for special values of the central charge, at which they become
local.

\bigskip \hrule \bigskip \centerline{\bf{References}}
\immediate\closeout\refs \vskip 0.5cm
  \message{References}\input references
\bye